\pdfoutput=1

\documentclass[3p,times,procedia]{elsarticle}
\usepackage{nupha_ecrc}

\pdfoutput=1
\volume{00}

\firstpage{1}

\journalname{Nuclear Physics A}

\runauth{Kathryn Meehan for the STAR Collaboration}

\jid{nupha}

\jnltitlelogo{Nuclear Physics A}

\usepackage{amssymb}

\biboptions{sort&compress}

\usepackage[figuresright]{rotating}

\usepackage{graphicx,wrapfig}
\usepackage[export]{adjustbox}
\usepackage{changepage}
\usepackage{subcaption}
\usepackage{xfrac}

\usepackage{placeins}
\let\Oldsection\section
\renewcommand{\section}{\FloatBarrier\Oldsection}
\let\Oldsubsection\subsection
\renewcommand{\subsection}{\FloatBarrier\Oldsubsection}
\let\Oldsubsubsection\subsubsection
\renewcommand{\subsubsection}{\FloatBarrier\Oldsubsubsection}

\begin{document}
\begin{frontmatter}

\dochead{XXVIth International Conference on Ultrarelativistic Nucleus-Nucleus Collisions\\ (Quark Matter 2017)}

\title{STAR Results from Au + Au Fixed-Target Collisions \newline at $\sqrt{s_{NN}}$ = 4.5 GeV}

\author{Kathryn Meehan for the STAR Collaboration\fnref{fn1}}
\fntext[fn1]{A list of members of the STAR Collaboration and acknowledgements can be found at the end of this issue.}

\address{UC Davis, One Shields Avenue, Davis, CA 94530}

\begin{abstract}
We present results from STAR's first dedicated fixed-target run conducted in 2015 with Au + Au collisions at $\sqrt{s_{NN}}$ \nolinebreak= 4.5 GeV. Directed flow of protons and lambdas, elliptic flow of identified hadrons, HBT radii, as well as pion, $K^{0}_{s}$, and $\Lambda$ spectra are compared with previous results from the AGS. These results demonstrate that STAR has good event reconstruction and particle identification capabilities in the fixed-target configuration. The implications of these results on future STAR fixed-target runs are discussed. 
\end{abstract}


\end{frontmatter}



\section{Introduction}

\begin{wrapfigure}[14]{r}{0.45\textwidth}
\centering
\vspace{-42pt}
\includegraphics[width=0.45\textwidth]{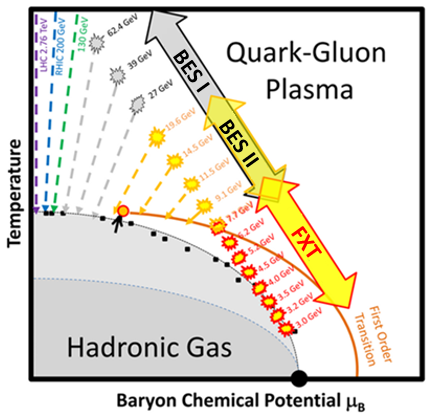}
\vspace{-20pt}
\caption{\label{fig:pd} QCD phase diagram showing the ranges of the BES
programs and the FXT program. The critical point and phase
transition lines are currently conjecture.}
\vspace{-20pt}
\end{wrapfigure}

At the top Au$+$Au RHIC energy of $\sqrt{s_{NN}}$ = 200 GeV, STAR has observed a new state of matter formed where partonic interactions dominate, referred to as the quark gluon plasma (QGP) \cite{starwhite}. The RHIC Beam Energy Scan (BES) Program was launched to explore the QCD phase diagram by tuning the collider to several different center-of-mass energies between 200 GeV and 7.7 GeV, shown in Figure \ref{fig:pd}. The goals of the BES program are to look for the turn-off of signatures of the quark gluon plasma (QGP), search for a possible QCD critical point, and study the nature of the phase transition between hadronic and partonic matter \cite{bes1white}. The results from the NA49 experiment at CERN have been used to suggest that the onset of deconfinement might occur at $\sqrt{s_{NN}} \approx$ 7 GeV, the low end of the BES range \cite{onsetNA49}. Additionally, studies of several interesting observables during the BES, including d$v_{1}$/dy of protons and lambdas, and net-proton higher moments, show interesting behavior below 20 GeV and could suggest a transition to a hadron-dominated regime \nolinebreak \cite{v1netpro,v2chargedpcle,Kurtosisnetpro}.  Data from energies lower than 7 GeV could help determine whether these behaviors are indicative of phase transitions or criticality. Furthermore, by going to energies below 7 GeV, where we expect no QGP formation, we can take control measurements of these signatures. However, it is not feasible to go to energies lower than 7 GeV in ``collider mode" because the luminosity of the beams decreases dramatically. Taking data at such low luminosities would require an impractical amount of time. By inserting a target into the beam pipe and only circulating one beam in the collider, we can instead study fixed-target collisions that avoid this luminosity challenge.

\vspace{20pt}
\begin{figure}[h]
\begin{center}
  \begin{subfigure}[b]{0.45\textwidth}
	\includegraphics[width=20pc]{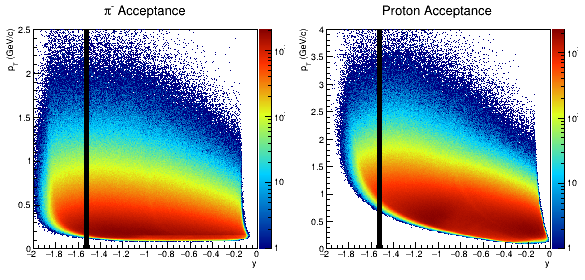}
    \caption{Pion and proton acceptance maps. The black line indicates midrapidity.}
    \label{fig:accept}
  \end{subfigure}
  \hspace{60pt}
  \begin{subfigure}[b]{0.3\textwidth}
  	\vspace{-20pt}
	\includegraphics[width=12pc]{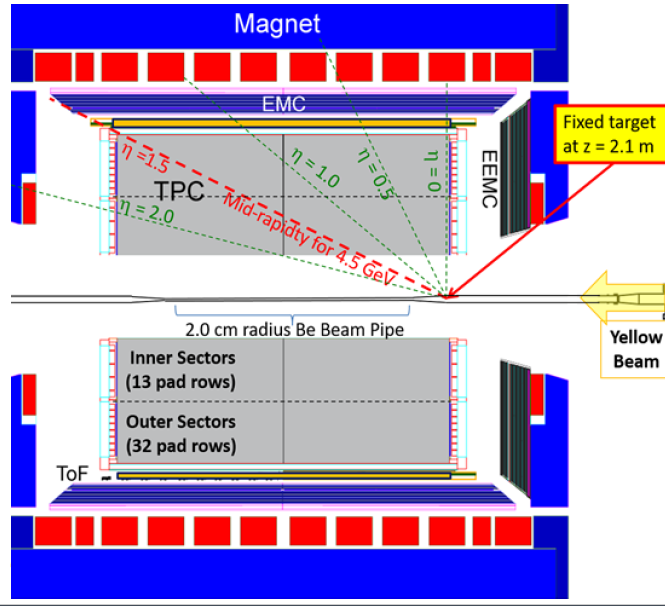}
    \caption{\label{fig:target} Diagram of the side-view of STAR.}
\end{subfigure}
\caption{Experiment set-up and acceptance.}
\label{fig:expt}
\end{center}
\end{figure}

In 2015 STAR conducted such a Au + Au $\sqrt{s_{NN}}$ = 4.5 GeV fixed-target test run to demonstrate STAR's capabilities in a fixed-target configuration. One beam was circulated in the collider and lowered to directly graze the edge of a ~1 mm thick (4\% interaction probability) gold foil target. As seen in Figure \ref{fig:target}, the target was placed at the edge of the TPC, about 211 cm away from the center of the detector to make use of the full tracking volume of the TPC. Approximately 1.3 million events were collected with a top $\sim$30\% centrality trigger. Pion and proton acceptance maps for this energy are shown in Figure \ref{fig:accept}.
\vspace{-5pt}

\section{Results}

\subsection{Comparison of STAR FXT and AGS Particle Yield Results}

\begin{figure}[h]

\begin{center}
\hspace{-35pt}
  \begin{subfigure}[b]{0.45\textwidth}
    \includegraphics[width=0.9\textwidth]{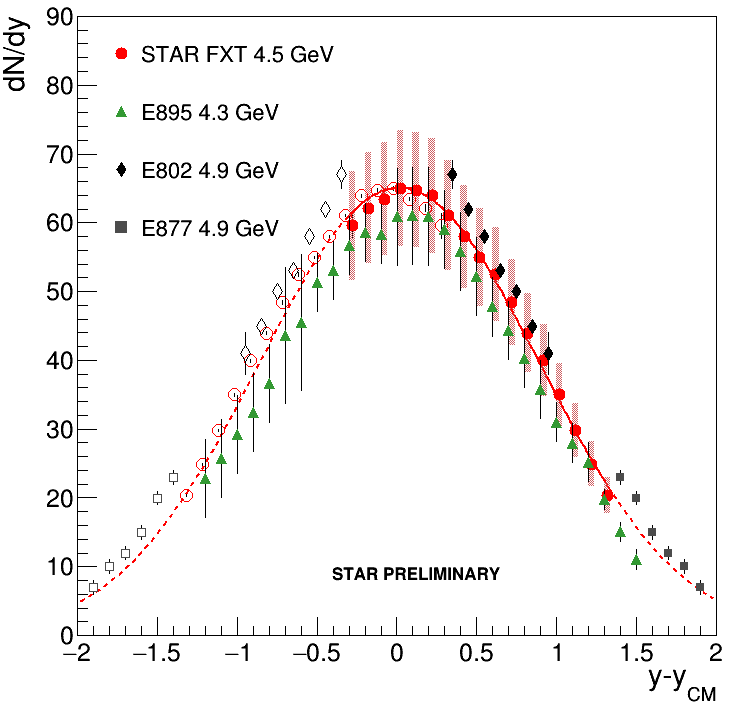}
    \caption{$\pi^{-}$ rapidity density distribution. STAR FXT data are plotted in red circles, while the E895, E892, and E877 AGS results are shown with other point styles. STAR FXT and E895 errors include systematic uncertainty. Open symbols are reflected. The red line is a Gaussian fit.}
    \label{fig:pimdNdy}
  \end{subfigure}
  \hspace{20pt}
  \begin{subfigure}[b]{0.45\textwidth}
    \hspace{-30pt}
    \includegraphics[width=1.2\textwidth]{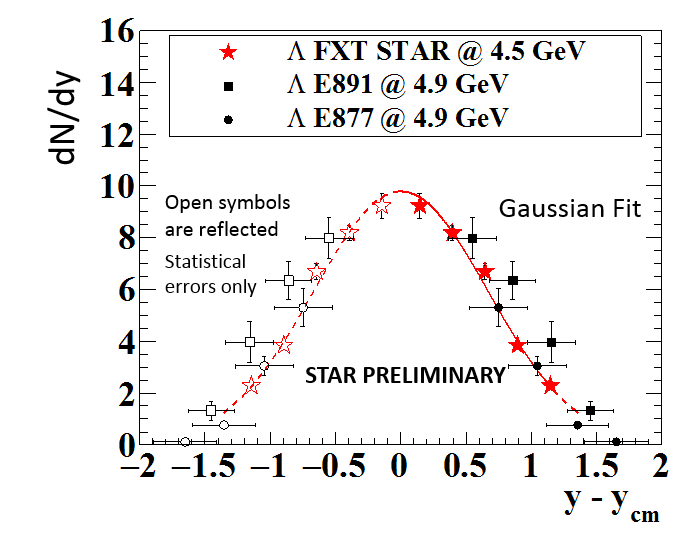}
    \vspace{-20pt}
    \caption{Lambda rapidity density. STAR FXT data are plotted as red stars. E891 and E877 points are plotted for comparison in black.}
    \vspace{30pt}
    \label{fig:lambdaRap}
  \end{subfigure}
  \hspace{-30pt}
\vspace{-5pt}
\caption{Rapidity density distributions.}
\label{fig:rap}
\vspace{-20pt}
\end{center}
\end{figure}

Spectra for $\pi^{-}$ for the top 5\% most central events were measured as a function of $m_{T}-m_{0}$ for several slices of rapidity width $\Delta$y = 0.1. These spectra were fit with Bose-Einstein functions to extract the rapidity density shown in Figure \ref{fig:pimdNdy}. The large systematic errors are due to the difference between using a Bose-Einstein fit and a Maxwell-Boltzmann fit to extrapolate the yield. The AGS results are also plotted for comparison. Within these systematics, the STAR FXT and E895 results, which are at a similar energy, are in good agreement \cite{KlayPions}. The E802 and E877 results at a slightly higher energy and plotted with just statistical uncertainties are also very close to the STAR FXT results. \cite{e802pim11.6, e877pim10.8}. 

Additionally, $\Lambda$ spectra were measured for the same centrality for several rapidity slices of width $\Delta$y=0.25 and the corresponding rapidity density distribution was obtained. The rapidity density was extracted using Maxwell-Boltzmann fits to the spectra. The $\Lambda$ rapidity density is shown in Figure \ref{fig:lambdaRap}.

The STAR FXT $\Lambda$ rapidity density (red) is plotted along with those measured by E877 and E891 \cite{e877lambda,e891lambda}. These AGS yields are in close agreement with our measurement, although slightly higher as expected due to their slightly higher energy. Overall, the amplitudes and widths of the $\pi^{-}$ and $\Lambda$ rapidity density distributions measured by STAR FXT are comparable with the AGS results.

\subsection{Comparison of STAR FXT and AGS Flow Results}

\begin{figure}[h]
\begin{center}
  \begin{subfigure}[b]{0.47\textwidth}
    \includegraphics[width=0.86\textwidth]{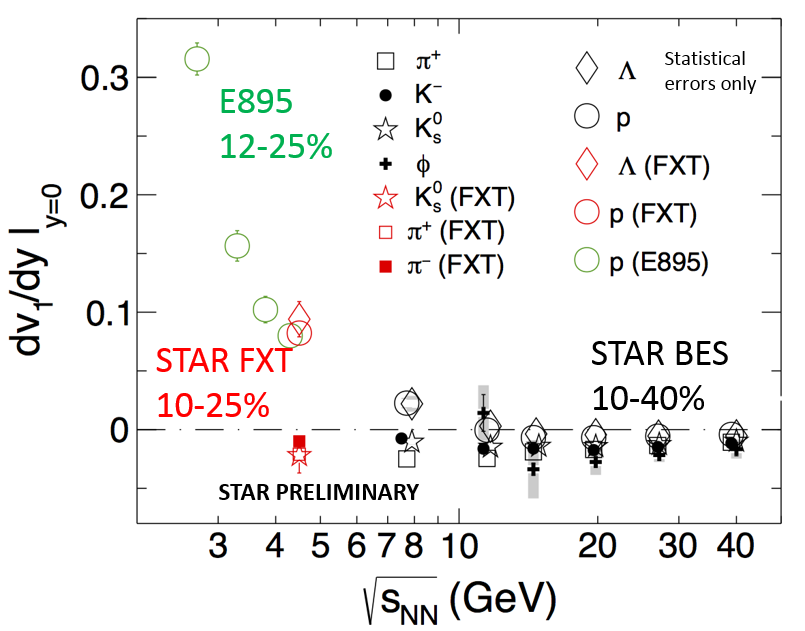}
    \caption{\label{fig:v1Trend} Energy dependence of dv$_{1}$/dy for several particle species measured by several experiments. STAR FXT measurements are shown in red. The STAR FXT $\pi^{+}$ and $\pi^{-}$ results are overlapping.}
  \end{subfigure}
  \hspace{10pt}
  \begin{subfigure}[b]{0.47\textwidth}
     \includegraphics[width=1.1\textwidth]{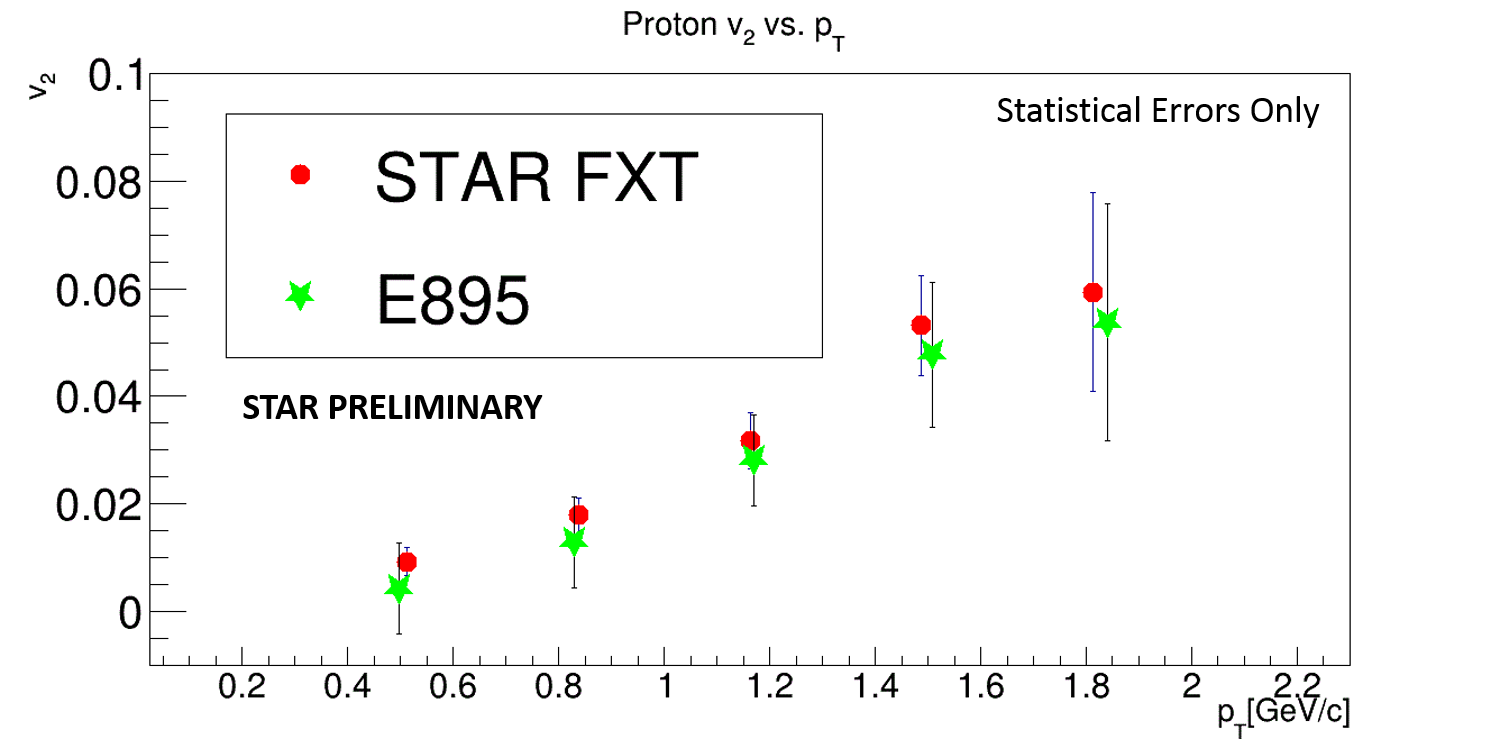}
    \caption{Measured v$_{2}$ of protons. Red points are STAR FXT data and green points are E895 data.}
    \label{fig:v2Expts}
    \vspace{20pt}
  \end{subfigure}
\caption{Directed and elliptic flow measurements.}  
\label{fig:flow}
\end{center}
\end{figure}

The directed flow ($\frac{dv_{1}}{dy}$) of pions, protons, K$^{0}_{s}$, and $\Lambda$ particles are also measured. In Figure \ref{fig:v1Trend}, STAR FXT measurements are plotted in red alongside measurements for several species and energies from both STAR BES I and E895 \cite{v1netpro,beslambdav1,e895v1}. The STAR FXT $\pi$ and K$^{0}_{s}$ measurements are negative and continue the trend for mesons observed during the STAR BES-I, down to lower energies. Similarly, the STAR FXT measurement for protons is in good agreement with the E895 result and both the proton and lambda measurement agree with a rising directed flow signal for baryons at lower energies. Once again, the STAR FXT measurements are consistent with previous experiments. 

Figure \ref{fig:v2Expts} displays our measurement (red) of proton elliptic flow ($v_{2}$) as a function of $p_{T}$ alongside the E895 measurement (green) \cite{e895v2}. It is clear the STAR FXT results successfully reproduce the E895 results.

\newpage

\subsection{Comparison of STAR FXT and AGS HBT Radii Results}

\begin{wrapfigure}[11]{r}{0.5\textwidth}
\centering
\vspace{-10pt}
	\includegraphics[width=0.4\textwidth]{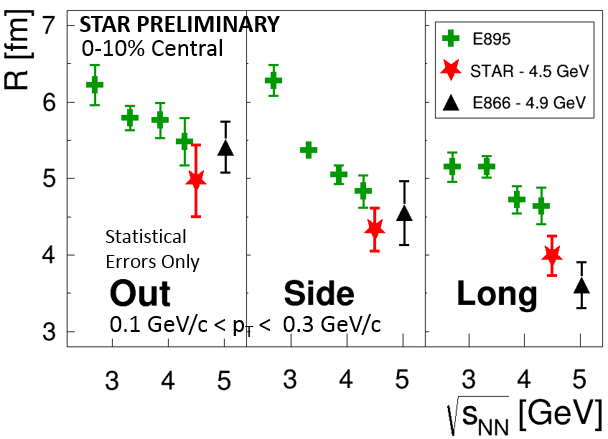}
\vspace{-10pt}
\caption{\label{fig:hbt}Measured HBT radii. The red stars are STAR FXT data, the green crosses are E895 data, and the black triangles are E866 data.}
\end{wrapfigure}

The HBT radii R$_{out}$, R$_{side}$, and R$_{long}$ were measured and are plotted in red as a function of energy in Figure \ref{fig:hbt}. The STAR FXT results are compared with E895 and E866 and are consistent with the energy dependence trend of these other experiments within uncertainties \cite{e895hbt,e866hbt}. 

\subsection{Dynamical Fluctuation Results}
\begin{wrapfigure}[12]{r}{0.5\textwidth}
\centering
\vspace{-55pt}
	\includegraphics[width=0.5\textwidth]{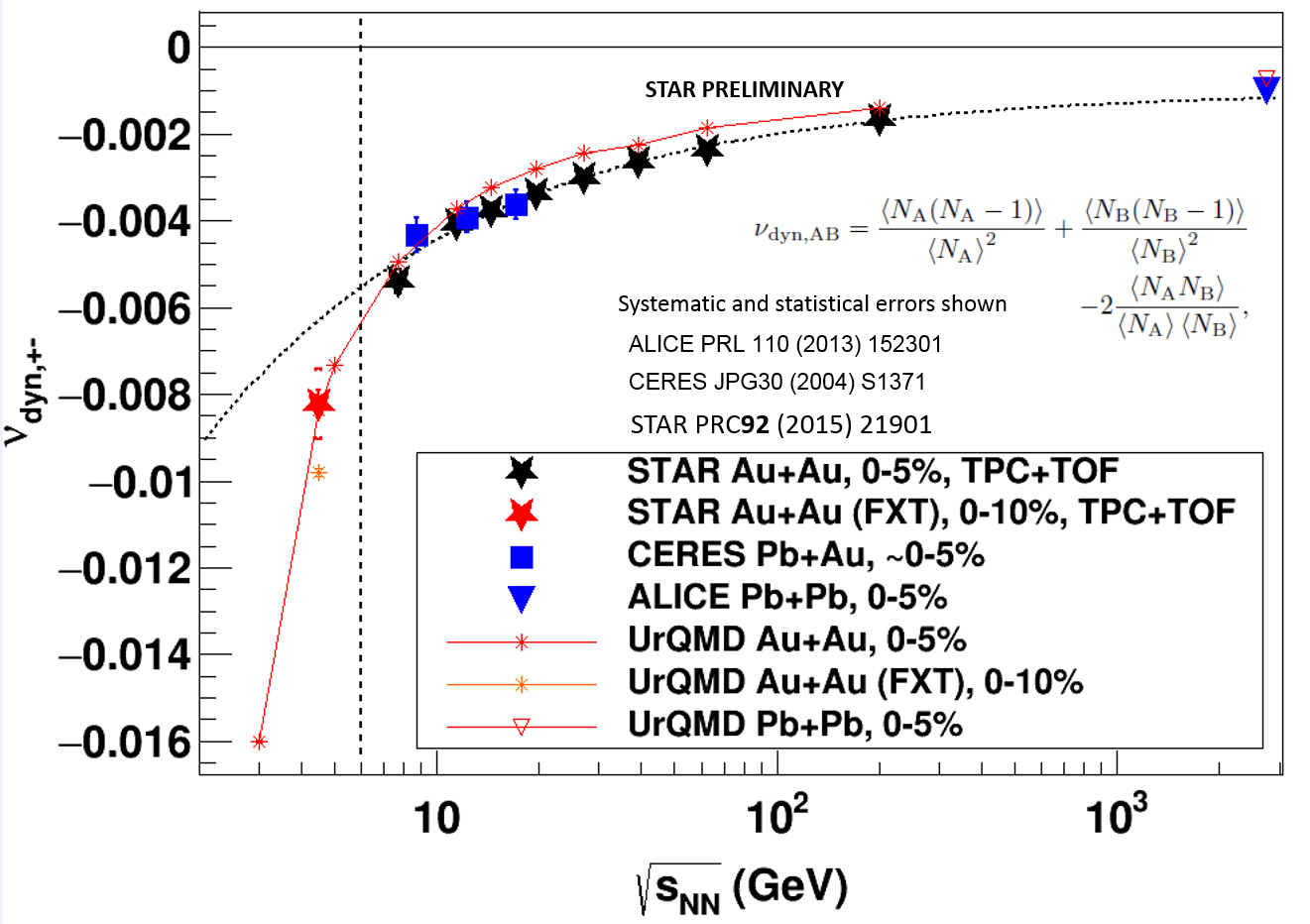}
\vspace{-20pt}
\caption{\label{fig:vdyn}$\nu_{dyn,+-}$ for a wide span of energies measured by STAR FXT, STAR BES-I, CERES, and ALICE. \cite{dynfbes,dynfceres,dynfalice}. The corresponding UrQMD simulations are drawn as lines.}
\end{wrapfigure}

The STAR FXT dynamical fluctuation result is plotted in Figure \ref{fig:vdyn}. This variable is defined in \cite{dynftheory}. The STAR FXT result continues the trend of increasing $\nu_{dyn,+-}$ with energy and is consistent with UrQMD simulations.

\section{Conclusions \& Future Plans} 
The results above demonstrate that in fixed-target mode the STAR detector is capable of measuring a wide array of physics results that are consistent with previous experiments. In the future we plan to do our main physics runs during STAR's Beam Energy Scan II program. Running for two days at each of our desired energies (shown in Figure \ref{fig:pd}) will allow us to collect up to 100 million events per energy and will extend the BES-II reach to about $\mu_{B}$=750 MeV. Furthermore, with the help of detector upgrades \cite{upgradesQM17}, the event plane resolution will be improved allowing better flow measurements. These upgrades will also improve acceptance and PID capabilities and will be critical to extending the fixed-target program up to $\sqrt{s_{NN}}$ = 7.7 GeV, which provides an ``overlap" energy with collider-mode analyses, enabling us to better understand acceptance effects. 

\bibliographystyle{elsarticle-num}
\bibliography{ms.bib}

\end{document}